\documentstyle[12pt,epsfig]{article}

\def\be{\begin{equation}}
\def\ee{\end{equation}}
\def\ba{\begin{eqnarray}}
\def\ea{\end{eqnarray}}

\begin{document}

\begin{titlepage}

\renewcommand{\thefootnote}{\fnsymbol{footnote}}

%\hfill SOGANG-MP 5/13

\vspace{0.3cm}

\begin{center}
{\Large\bf Local free-fall Temperature \\of GMGHS Black Holes}
\end{center}

\begin{center}
Yong-Wan Kim\footnote{Electronic address:
ywkim65@gmail.com}$^{1}$, Jaedong Choi\footnote{Electronic
address: choijdong@gmail.com}$^{2}$, Young-Jai
Park$\footnote{Electronic address: yjpark@sogang.ac.kr}^{1,3}$\par
\end{center}

\begin{center}

{{${}^{1}$Center for Quantum Spacetime, Sogang University, Seoul
121-742, Korea, \\${}^{2}$Department of Basic Science, Korea Air
Force Academy,}\par {Ssangsu, Namil, Cheongwon, Chungbuk 363-849,
Korea,}\par } {${}^{3}$Department of Physics, Sogang University,
Seoul 121-742, Korea}\par
\end{center}

\vskip 0.5cm
\begin{center}
{\today}
\end{center}

\vfill

\begin{abstract}
We obtain a (5+1)-dimensional global flat embedding of the
Gibbons-Maeda-Garfinkle-Horowitz-Strominger spacetime in Einstein
frame, and a (5+2)-dimensional global flat embedding in string
frame. We show that the local free-fall temperatures for freely
falling observers in each frames are finite at the event horizons,
while the local temperatures for fiducial observers are divergent.
We also observe that the local free-fall temperatures differ
between the two frames.
\end{abstract}

\vskip20pt

PACS numbers: 04.70.Dy, 04.20.Jb, 04.62.+v
\vskip15pt
Keywords: GMGHS spacetime, Global flat embedding, Unruh effect \\
\end{titlepage}

\newpage

%%%%%%%%%%%%%%%%%%%%%%%%%%%%%%%%%%%%%%%%%%%%%%%%%%%%%%%%%%%%%%%%%%%%%%%
\section{ Introduction}
\setcounter{equation}{0}
\renewcommand{\theequation}{\arabic{section}.\arabic{equation}}
%%%%%%%%%%%%%%%%%%%%%%%%%%%%%%%%%%%%%%%%%%%%%%%%%%%%%%%%%%%%%%%%%%%%%%

Hawking discovered that a black hole emits thermal radiation with
characteristic Hawking temperature $T_H$ as seen by asymptotic
observers~\cite{Hawking:1974sw}. The local temperature measured by a
fiducial observer at a finite distance from a black hole is then
described by the Tolman temperature~\cite{Tolman:1930zza}
 \be
 T_{\rm FID}=\frac{T_H}{\sqrt{-g_{\mu\nu}\xi^\mu\xi^\nu}},
 \ee
where $\xi^\mu$ is a timelike Killing vector. Later,
Unruh~\cite{Unruh:1976db} showed that a uniformly accelerated
observer in flat spacetime, with proper acceleration $a$, will
detect thermal radiation at the Unruh temperature
 \be
  T_U=\frac{a}{2\pi}.
 \ee
These two effects are related, {\it i.e.,} the Hawking effect for a
fiducial observer in a black hole spacetime can be considered as the
Unruh effect for a uniformly accelerated observer in a higher
dimensional global embedding Minkowskian spacetime (GEMS). After
confirming these ideas in an analysis of de Sitter
(dS)~\cite{Narnhofer:1996zk} and anti-de Sitter (AdS)
spacetimes~\cite{Deser:1997ri} and their corresponding GEMSs, Deser
and Levin~\cite{Deser:1998xb} have shown that the GEMS approach
provides a unified derivation of temperature for
Ba\~nados-Teitelboim-Zanelli, Schwarzschild-AdS(-dS), and
Reissner-Nordstr\"om (RN) spacetimes. These results have since been
extended to a wide variety of curved
spacetimes~\cite{Kim:2000ct,Hong:2000kn,Hong:2003xz,Chen:2004qw,Santos:2004ws,Banerjee:2010ma,Cai:2010bv,Majhi:2011yi,Hu:2011yx},
and also see references therein. Recently, Brynjolfsson and
Thorlacius~\cite{Brynjolfsson:2008uc} have used the GEMS approach to
define a local temperature for a freely falling observer outside
Schwarzschild(-AdS) and RN spacetimes, and we have extended the
results to RN-AdS spacetime~\cite{Kim:2009ha}. In particular, here a
local free-fall temperature is defined at special turning points of
radial geodesics where a freely falling observer is momentarily at
rest with respect to a black hole. It was shown that the local
free-fall temperature approaches the Hawking temperature at spatial
infinity, while it remains finite at the event horizon.

On the other hand, the spherically symmetric static charged black
hole solution in low energy effective theory of heterotic string
theory in four dimension was found by Gibbons,
Maeda~\cite{Gibbons:1987ps}, and independently, by Garfinkle,
Horowitz, Strominger~\cite{Garfinkle:1990qj}, by turning
antisymmetric tensor gauge fields off, from now on refereed to as
the Gibbons-Maeda-Garfinkle-Horowitz-Strominger (GMGHS) solution.
See
also~\cite{Horowitz:1992jp,Rakhmanov:1993yd,Chan:1996sx,Bose:1998yp}.
After these works, there were enormous interests in the spherically
symmetric static charged black
holes~\cite{Natsuume:1994hd,Galtsov:1995va,Xulu:1997fk,Shu:2004fj,Sur:2005pm,Sheykhi:2007wg,Jiang:2008zzg,Chen:2010kn,Fernando:2011ki,Choi:2013bec}
and see also references therein. In particular, this GMGHS black
hole spacetime can be described by the solutions either in Einstein
or string frames. In Einstein frame, the action is in the form of
the Einstein-Hilbert action, while in string frame strings directly
couple to the metric of $e^{2\phi}g_{\mu\nu}$ where $e^{2\phi}$ is a
conformal factor and $g_{\mu\nu}$ is the Einstein metric. Even
though the solutions in the two frames are related by a conformal
transformation so that they are mathematically isomorphic to each
other~\cite{Faraoni:1998qx}, however, there are differences in some
of the physical properties of the black hole solutions in the two
frames~\cite{Casadio:1998wu}.

In this paper, we wish to study the GEMSs of the GMGHS black hole
spacetimes both in the Einstein and string frames, and investigate
how different GEMS embeddings are in the two frames. The GEMS
approach is also important on its own, since it gives a powerful
tool that simplifies the study of black hole physics by working
instead, but equivalently, in an accelerated frame in a higher
dimensional flat spacetime. In section 2, we briefly review the
structure of the GEMS of the curved Schwarzschild and RN spacetimes,
and their local free-fall temperatures. In section 3, we apply this
GEMS approach to the charged GMGHS black hole spacetimes in the
Einstein frame. In section 4, we also embed both the magnetically
and electrically charged GMGHS black hole spacetimes in the string
frame into higher-dimensional flat spacetimes, and find their local
free-fall temperatures. Our conclusions are drawn in section 5.

%%%%%%%%%%%%%%%%%%%%%%%%%%%%%%%%%%%%%%%%%%%%%%%%%%%%%%%%%%%%%%%%%%%%%%%
\section{Schwarzschild and RN spacetimes}
\setcounter{equation}{0}
\renewcommand{\theequation}{\arabic{section}.\arabic{equation}}
%%%%%%%%%%%%%%%%%%%%%%%%%%%%%%%%%%%%%%%%%%%%%%%%%%%%%%%%%%%%%%%%%%%%%%%

%%%%%%%%%%%%%%%%%%%%%%%%%%%%%%%%%%%%%%%%%%%%%%%%%%%%%%%%%%%%%%%%%%%%%%
\subsection{Schwarzschild spacetime}
%%%%%%%%%%%%%%%%%%%%%%%%%%%%%%%%%%%%%%%%%%%%%%%%%%%%%%%%%%%%%%%%%%%%%%

The Schwarzschild spacetime described by the metric
 \be
 ds^2=-\left(1-\frac{2M}{r}\right)dt^2+\left(1-\frac{2M}{r}\right)^{-1}dr^2
      +r^2(d\theta^2+\sin^2\theta d\phi^2)
 \ee
can be embedded into a (5+1)-dimensional Minkowskian spacetime
 \be
 ds^2=\eta_{IJ}z^I z^J
 \ee
with a metric $\eta_{IJ}={\rm diag}(-1,1,1,1,1,1)$. Explicitly, the
(5+1)-dimensional Minkowskian spacetime is given by the
transformations~\cite{Fronsdal:1959zza}
 \ba\label{schemb1}
 z^0&=&\frac{1}{k_H}\sqrt{1-\frac{2M}{r}}\sinh(k_H t),\\
 z^1&=&\frac{1}{k_H}\sqrt{1-\frac{2M}{r}}\cosh(k_H t),\\
 z^2&=&\int dr \sqrt{\frac{2M(r^2+2M r+4M^2)}{r^3}},\\
 z^3&=&r\sin\theta\cos\phi,\\
 z^4&=&r\sin\theta\sin\phi,\\
 \label{schemb2}
 z^5&=&r\cos\theta.
 \ea
Note that $k_H(=\frac{1}{4M})$ is the surface gravity and the event
horizon $r_H$ is $2M$. An observer, who is uniformly accelerated in
the (5+1)-dimensional flat spacetime, follows a hyperbolic
trajectory described by proper acceleration
 \be\label{Sch-accel}
 a^{-2}_6=(z^1)^2-(z^0)^2=16M^2\left(1-\frac{2M}{r}\right).
 \ee
Thus, as was shown by Unruh~\cite{Unruh:1976db}, the Unruh
temperature can be read as
 \be\label{unruh-sch}
 T_U=\frac{a_6}{2\pi}=\frac{1}{8\pi M\sqrt{1-\frac{2M}{r}}}.
 \ee
In fact, this corresponds to the local temperature measured by a
fiducial observer at a finite distance from the black hole, also
called the fiducial temperature
 \be\label{fid-temp}
 T_{\rm FID}=\frac{T_H}{\sqrt{-g_{00}}},
 \ee
where the Hawking temperature $T_H$ measured by an asymptotic
observer is
 \be
 T_H=\frac{1}{8\pi M}.
 \ee
As results, the Hawking effect for a fiducial observer in the black
hole spacetime can be said to be the Unruh effect for a uniformly
accelerated observer in a higher-dimensional flat spacetime.

Now, consider a freely falling observer who is dropped from rest at
$r=r_0$ and at $\tau=0$. For a freely falling observer, there are
turning points~\cite{Brynjolfsson:2008uc} of radial geodesics where
a freely falling observer is momentarily at rest with respect to
black holes. Making use of the orbit equations~\cite{Kim:2009ha}
 \be\label{orbiteq}
 \frac{dt}{d\tau}=\frac{\sqrt{f(r_0)}}{f(r)},~~~
 \frac{dr}{d\tau}=-\sqrt{f(r_0)-f(r)},~~{\rm
 with}~f(r)=1-\frac{2M}{r},
 \ee
one can obtain the $\widetilde{a}_6$ acceleration given by
 \be\label{a6-sch}
 (\widetilde{a}_6)^2=\frac{r^3+2Mr^2+4M^2r+8M^3}{16M^2r^3}.
 \ee
Taking the local free-fall temperature
 \be
 T_{\rm FFAR}=\frac{\widetilde{a}_6}{2\pi}
 \ee
measured by the freely falling observer at rest (FFAR) to be the
local Unruh temperature, one obtains
 \be\label{ffar-sch}
 T_{\rm FFAR}=\frac{1}{8\pi M}\sqrt{\frac{r^3+2Mr^2+4M^2r+8M^3}{r^3}},
 \ee
which is reduced to the Hawking temperature $T_H$ at infinity. It is
important to note that the local free-fall temperature at the event
horizon is finite as $T_{\rm FFAR}=2T_H$, while the local
temperature $T_{\rm FID}$ (\ref{fid-temp}) for the fiducial observer
diverges as $r\rightarrow r_H$.

%%%%%%%%%%%%%%%%%%%%%%%%%%%%%%%%%%%%%%%%%%%%%%%%%%%%%%%%%%%%%%%%%%%%%%
\subsection{RN spacetime}
%%%%%%%%%%%%%%%%%%%%%%%%%%%%%%%%%%%%%%%%%%%%%%%%%%%%%%%%%%%%%%%%%%%%%%

In order to embed the charged RN black hole given by
 \be
 ds^2=-\left(1-\frac{2M}{r}+\frac{Q^2}{r^2}\right)dt^2+\left(1-\frac{2M}{r}+\frac{Q^2}{r^2}\right)^{-1}dr^2
      +r^2(d\theta^2+\sin^2\theta d\phi^2),
 \ee
into the higher dimensional flat spacetime, one needs to introduce
one more time dimension, compared with the GEMS embedding of the
Schwarzschild spacetime. As a result, the embedded flat spacetime is
described by the (5+2)-dimensional Minkowskian spacetime
 \be
 ds^2=\eta_{IJ}z^I z^J
 \ee
with a metric $\eta_{IJ}={\rm diag}(-1,1,1,1,1,1,-1)$ given by the
transformations~\cite{Deser:1998xb,Kim:2000ct}
 \ba\label{embrn1}
 z^0&=&\frac{1}{k_H}\sqrt{1-\frac{2M}{r}+\frac{Q^2}{r^2}}\sinh(k_H t),\\
 z^1&=&\frac{1}{k_H}\sqrt{1-\frac{2M}{r}+\frac{Q^2}{r^2}}\cosh(k_H t),\\
 z^2&=&\int dr \sqrt{\frac{2M r^2+r^2_H(r+r_H)}{r^2(r-M+\sqrt{M^2-Q^2})}},\\
 z^3&=&r\sin\theta\cos\phi,\\
 z^4&=&r\sin\theta\sin\phi,\\
 z^5&=&r\cos\theta,\\
 \label{embrn2}
 z^6&=&\int dr\sqrt{\frac{Q^2r^4_H}{r^4(M^2-Q^2)}},
 \ea
where $r_H(=M+\sqrt{M^2-Q^2})$ denotes the outer horizon and the
surface gravity is $k_H=\sqrt{M^2-Q^2}/r^2_H$. Note that since
$z^6\rightarrow 0$ in the $Q\rightarrow 0$ limit, the
transformations are reduced to the Schwarzschild ones.

For an uniformly accelerating observer, the $a_7$-acceleration is
given by
 \be
 a^{-2}_7=\frac{r^4_H}{M^2-Q^2}\left(1-\frac{2M}{r}+\frac{Q^2}{r^2}\right),
 \ee
and thus the fiducial temperature corresponding to the Unruh one is
read as
 \be
 T_{\rm FID}=\frac{\sqrt{M^2-Q^2}}{2\pi r^2_H\sqrt{1-\frac{2M}{r}+\frac{Q^2}{r^2}}}.
 \ee
Then, one can obtain the Hawking temperature $T_H$ by taking
$r\rightarrow\infty$ as
 \be
  T_H=\frac{\sqrt{M^2-Q^2}}{2\pi r^2_H}.
 \ee
On the other hand, for the RN black hole, the local free-fall
temperature seen by a freely falling observer is given by
 \be
 T_{FFAR}=\frac{1}{2\pi r^2_H}\sqrt{\frac{(M^2-Q^2)(r^3+r_Hr^2+r^2_Hr+r^3_H)}{r^2(r-M+\sqrt{M^2-Q^2})}
              -\frac{Q^2r^4_H}{r^4}},
 \ee
which is reduced to the local free-fall temperature $T_{FFAR}$ of
the Schwarzschild black hole in Eq.~(\ref{ffar-sch}) in the $Q=0$
limit~\cite{Brynjolfsson:2008uc,Kim:2009ha}.

%%%%%%%%%%%%%%%%%%%%%%%%%%%%%%%%%%%%%%%%%%%%%%%%%%%%%%%%%%%%%%%%%%%%%%
\section{GMGHS spacetime in Einstein frame}
\setcounter{equation}{0}
\renewcommand{\theequation}{\arabic{section}.\arabic{equation}}
%%%%%%%%%%%%%%%%%%%%%%%%%%%%%%%%%%%%%%%%%%%%%%%%%%%%%%%%%%%%%%%%%%%%%%

Now, let us start with the GMGHS action~\cite{Garfinkle:1990qj} in
the Einstein frame
 \be\label{E-action}
 S=\int d^4x \sqrt{-g}\left(R-2(\nabla\phi)^2-e^{-2\phi}F_{\mu\nu}F^{\mu\nu}\right),
 \ee
where $R$ is the scalar curvature, $\phi$ is a dilaton, and
$F_{\mu\nu}$ is the Maxwell field. Spherically symmetric static
charged
solutions~\cite{Garfinkle:1990qj,Horowitz:1992jp,Rakhmanov:1993yd,Chan:1996sx,Bose:1998yp}
to equations of motion of the action~(\ref{E-action}) are given by
 \be\label{metric-e}
 ds^2=-\left(1-\frac{2M}{r}\right)dt^2+\left(1-\frac{2M}{r}\right)^{-1}dr^2
      +r\left(r-\frac{Q^2}{M}\right)(d\theta^2+\sin^2\theta
      d\phi^2),
 \ee
where $Q$ is related to the magnetic and electric charges defined by
 \be
 F_{\theta\phi}=Q\sin\theta,~~~F_{tr}=\frac{Qe^{4\phi}}{r^2},
 \ee
respectively. Moreover, they have the relation with the dilaton as
 \be
 e^{-2\phi}=1-\frac{Q^2}{Mr},~~~e^{-2\phi}=1+\frac{Q^2}{Mr}
 \ee
for the magnetically and electrically charged black holes,
respectively. This charged case seems to the string analog of the RN
spacetime.

However, this (3+1)-dimensional curved spacetime (\ref{metric-e})
can be embedded into a (5+1)-dimensional Minkowskian spacetime
 \be
 ds^2=\eta_{IJ}z^I z^J
 \ee
with a flat metric $\eta_{IJ}={\rm diag}(-1,1,1,1,1,1)$, similar to
the Schwarzschild spacetime in contrast to the RN case. Here, the
(5+1)-dimensional Minkowskian spacetime is explicitly given by the
transformations
 \ba
 z^0&=&\frac{1}{k_H}\sqrt{1-\frac{2M}{r}}\sinh(k_H t),\\
 z^1&=&\frac{1}{k_H}\sqrt{1-\frac{2M}{r}}\cosh(k_H t),\\
 z^2&=&\int dr \sqrt{\frac{2M(r^2+2M r+4M^2)}{r^3}
            +\frac{Q^2}{4M}\left(\frac{1}{r}-\frac{1}{r-\frac{Q^2}{M}}\right)},\\
 z^3&=&\sqrt{r\left(r-\frac{Q^2}{M}\right)}\sin\theta\cos\phi,\\
 z^4&=&\sqrt{r\left(r-\frac{Q^2}{M}\right)}\sin\theta\sin\phi,\\
 z^5&=&\sqrt{r\left(r-\frac{Q^2}{M}\right)}\cos\theta,
% z^6&=&\int dr \sqrt{\frac{Q^2}{4M}\left(\frac{1}{r-\frac{Q^2}{M}}\right)},
 \ea
which reduce to the transformations~(\ref{schemb1})-(\ref{schemb2})
in the $Q\rightarrow 0$ limit. However, they are very different from
the embeddings of the RN spacetime~(\ref{embrn1})-(\ref{embrn2}).
Note here that $k_H(=\frac{1}{4M})$ is the surface gravity and the
event horizon $r_H$ is given by $2M$ as like the Schwarzschild case.
However, the area of the two sphere of constant $t$ and $r$ is
smaller than the Schwarzschild one due to the presence of the
dilaton.

Now, for an uniformly accelerating observer, the $a_6$-acceleration
is given by
 \be
 a_6=\frac{1}{4M\sqrt{1-\frac{2M}{r}}}.
 \ee
Then, the fiducial temperature corresponding to the Unruh one is
simply written
 \be
 T_{\rm FID}=\frac{1}{8\pi M\sqrt{1-\frac{2M}{r}}}.
 \ee
As results, we see that these are the same with the ones of the
Schwarzschild spacetimes (\ref{unruh-sch}). Moreover, it is
independent of the charge $Q$.

On the other hand, making use of the orbit equations (\ref{orbiteq})
with $f(r)=1-\frac{2M}{r}$, we can obtain the $\widetilde{a}_6$
acceleration for a freely falling observer given by
 \be
 (\widetilde{a}_6)^2=\frac{r^3+2Mr^2+4M^2r+8M^3}{16M^2r^3},
 \ee
which is exactly the same with the 6-acceleration (\ref{a6-sch}) for
the freely falling observer in the Schwarzschild spacetime. As a
result, we have the same local Unruh temperature $T_{FFAR}$
(\ref{ffar-sch}) for the freely fall observer at rest.

In this respect, we know that even though the transformations of the
embedding coordinates are differently given due to the charge of the
GMGHS spacetime in the Einstein frame, it does not have any affects
on the accelerations and the corresponding Unruh/free-fall
temperatures.

%%%%%%%%%%%%%%%%%%%%%%%%%%%%%%%%%%%%%%%%%%%%%%%%%%%%%%%%%%%%%%%%%%%%%%
\section{GMGHS spacetime in string frame}
\setcounter{equation}{0}
\renewcommand{\theequation}{\arabic{section}.\arabic{equation}}
%%%%%%%%%%%%%%%%%%%%%%%%%%%%%%%%%%%%%%%%%%%%%%%%%%%%%%%%%%%%%%%%%%%%%%

In the string frame, the GMGHS action is described by
 \be\label{S-action}
  S=\int d^4x \sqrt{-g}e^{-2\phi}\left(R+4(\nabla\phi)^2-F_{\mu\nu}F^{\mu\nu}\right),
 \ee
which is related to the action~(\ref{E-action}) in the Einstein
frame through the conformal transformation of $g^{\rm
S}_{\mu\nu}=e^{2\phi}g^{\rm E}_{\mu\nu}$.

%%%%%%%%%%%%%%%%%%%%%%%%%%%%%%%%%%%%%%%%%%%%%%%%%%%%%%%%%%%%%%%%%%%%%%
\subsection{Magnetically charged solution}
%%%%%%%%%%%%%%%%%%%%%%%%%%%%%%%%%%%%%%%%%%%%%%%%%%%%%%%%%%%%%%%%%%%%%%

Now, let us study the magnetically charged GMGHS solution in the
string frame, which is given by
 \be\label{metric-s1}
 ds^2=-\frac{\left(1-\frac{2M}{r}\right)}{\left(1-\frac{Q^2}{Mr}\right)}dt^2
    +\frac{dr^2}{\left(1-\frac{2M}{r}\right)\left(1-\frac{Q^2}{Mr}\right)}
    +r^2(d\theta^2+\sin^2\theta d\phi^2).
 \ee
The (3+1)-dimensional curved spacetime can be embedded in a
(5+2)-dimensional Minkowskian spacetime
 \be
 ds^2=\eta_{IJ}z^I z^J
 \ee
with a metric $\eta_{IJ}={\rm diag}(-1,1,1,1,1,1,-1)$. The
transformations are
 \ba
 z^0&=&\frac{1}{k_H}\sqrt{\frac{1-\frac{2M}{r}}{1-\frac{Q^2}{Mr}}}\sinh(k_H t),\\
 z^1&=&\frac{1}{k_H}\sqrt{\frac{1-\frac{2M}{r}}{1-\frac{Q^2}{Mr}}}\cosh(k_H t),\\
 z^2&=&\int dr \sqrt{\frac{\frac{2M}{r}+\frac{4M^2}{r^2}+\frac{8M^3}{r^3}
                     +\frac{Q^2}{Mr}+\frac{2Q^4}{Mr^3}+\frac{Q^6}{M^3r^3}}{(1-\frac{Q^2}{Mr})^3}},\\
 z^3&=&r\sin\theta\cos\phi,\\
 z^4&=&r\sin\theta\sin\phi,\\
 z^5&=&r\cos\theta,\\
 z^6&=&\int dr \sqrt{\frac{\frac{4Q^2}{r^2}(1+\frac{2M}{r}+\frac{Q^2}{2M^2})}{(1-\frac{Q^2}{Mr})^3}}.
 \ea
Changing the frame from the Einstein to the string makes different
the embeddings of the GMGHS spacetimes. Note that
$k_H(=\frac{1}{4M})$ is the surface gravity and the event horizon
$r^m_H$ is also $2M$ as like the Schwarzschild case.

The $a_7$-acceleration for an uniformly accelerating observer is
given by
 \be
 a_7=\frac{1}{4M}\sqrt{\frac{1-\frac{Q^2}{Mr}}{1-\frac{2M}{r}}}.
 \ee
Then, the fiducial temperature is given by
 \be\label{unruh-m}
 T_{\rm FID}=\frac{1}{8\pi M}\sqrt{\frac{1-\frac{Q^2}{Mr}}{1-\frac{2M}{r}}}.
 \ee
This becomes the Hawking temperature $T_H$ at the asymptotic
infinity, as expected.

On the other hand, for the case of the magnetically charged
solution (\ref{metric-s1}), we have orbit equations as
 \ba
 \frac{dt}{d\tau}&=&  \left(\frac{1-\frac{2M}{r_0}}{1-\frac{Q^2}{Mr_0}}\right)^{1/2}
                     \left(\frac{1-\frac{Q^2}{Mr}}{1-\frac{2M}{r}}\right),
                    \nonumber\\
 \frac{dr}{d\tau}&=&-\left[-\left(1-\frac{2M}{r}\right)\left(1-\frac{Q^2}{Mr}\right)+\left(1-\frac{Q^2}{Mr}\right)^2
                    \left(\frac{1-\frac{2M}{r_0}}{1-\frac{Q^2}{Mr_0}}\right)\right]^{1/2},\nonumber\\
 \ea
which can be used to get the $\widetilde{a}_7$ acceleration for a
freely falling observer given by
 \ba\label{local-a7-mag}
 (\widetilde{a}_7)^2=\frac{1+\frac{2M}{r}+\frac{4M^2}{r^2}+\frac{8M^3}{r^3}
                       -\frac{2Q^2}{Mr}\left(1+\frac{2M}{r}+\frac{4M^2}{r^2}\right)
                       +\frac{Q^4}{M^2r^2}\left(1+\frac{2M}{r}\right)}
                        {16M^2\left(1-\frac{Q^2}{Mr}\right)}.
 \ea
As a result, we have the local free-fall temperature for the freely
falling observer at rest as
 \be\label{ffar-mag}
 T_{FFAR}=\frac{\widetilde{a}_7}{2\pi}.
 \ee
At the asymptotic infinity, the local free-fall temperature
becomes the Hawking temperature $T_H$. Note also that the local
free-fall temperature in the string frame depends on the charge,
while the local free-fall temperature in the Einstein frame does
not. Moreover, when $Q\rightarrow 0$, the temperature
(\ref{ffar-mag}) reduces to the local free-fall temperature
(\ref{ffar-sch}) for the Schwarzschild spacetime. At the event
horizon, the local free-fall temperature remains finite as
 \be\label{srootdiv}
  T_{FFAR}=\frac{\sqrt{1-\frac{Q^2}{2M^2}}}{4\pi M},
 \ee
while the fiducial temperature (\ref{unruh-m}) diverges. It also
depends on the charge, and has lower temperature, compared with the
temperatures of $T_{\rm FFAR}=2T_H$ in the Einstein frame and for
the Schwarzschild spacetime.

On the other hand, in the extremal limit of $2M^2=Q^2$, the local
free-fall temperature becomes
 \be\label{extT-mag}
 T_{FFAR}=\frac{1}{8\pi M}
 \ee
so that the local free-fall temperature for the extremal black hole
appears to be constant, depending on the mass of the black hole, for
the freely falling observer.

%%%%%%%%%%%%%%%%%%%%%%%%%%%%%%%%%%%%%%%%%%%%%%%%%%%%%%%%%%%%%%%%%%%%%%
\subsection{Electrically charged solution}
%%%%%%%%%%%%%%%%%%%%%%%%%%%%%%%%%%%%%%%%%%%%%%%%%%%%%%%%%%%%%%%%%%%%%%

The electrically charged GMGHS solution in the string frame is
given by
 \be\label{metric-s2}
 ds^2=-\frac{\left(1+\frac{Q^2-2M^2}{Mr}\right)}{\left(1+\frac{Q^2}{Mr}\right)^2}dt^2
    +\frac{dr^2}{\left(1+\frac{Q^2-2M^2}{Mr}\right)}
    +r^2(d\theta^2+\sin^2\theta d\phi^2).
 \ee
This spacetime can be also embedded in a (5+2)-dimensional
Minkowskian spacetime
 \be
 ds^2=\eta_{IJ}z^I z^J
 \ee
with a metric $\eta_{IJ}={\rm diag}(-1,1,1,1,1,1,-1)$. The
transformations are
 \ba
 z^0&=&\frac{1}{k_H}\frac{\sqrt{1+\frac{Q^2-2M^2}{Mr}}}{\left(1+\frac{Q^2}{Mr}\right)}\sinh(k_H t),\\
 z^1&=&\frac{1}{k_H}\frac{\sqrt{1+\frac{Q^2-2M^2}{Mr}}}{\left(1+\frac{Q^2}{Mr}\right)}\cosh(k_H t),\\
 z^2&=&\int dr  \frac{\sqrt{\frac{2M}{r}\left[1+\frac{2M}{r}+\frac{4M^2}{r^2}
          +\frac{Q^2}{Mr}\left(2+\frac{2M}{r}+\frac{Q^2}{Mr}+\frac{4MQ^2}{r^3}\right)\right]}}
                    {\left(1+\frac{Q^2}{Mr}\right)^2},\\
 z^3&=&r\sin\theta\cos\phi,\\
 z^4&=&r\sin\theta\sin\phi,\\
 z^5&=&r\cos\theta,\\
 z^6&=&\int dr  \frac{\sqrt{\frac{Q^2}{Mr}\left[1+\frac{16M^3}{r^3}
                     +\frac{Q^2}{Mr}\left(3+\frac{4M^2}{r^2}+\frac{3Q^2}{Mr}
                     +\frac{4MQ^2}{r^3}+\frac{Q^4}{M^2r^2}\right)\right]}}
                    {\left(1+\frac{Q^2}{Mr}\right)^2}.\nonumber\\
 \ea
Note that $k_H(=\frac{1}{4M})$ is the surface gravity and the event
horizon $r^e_H$ is given by $2M-Q^2/M$, which is shifted from $2M$
by $Q^2/M$ as compared with the event horizon $r^m_H$ of the
magnetically charged solution.

The $a_7$-acceleration for an uniformly accelerating observer is
given by
 \be
 a_7=\frac{1+\frac{Q^2}{Mr}}{4M\sqrt{1+\frac{Q^2-2M^2}{Mr}}},
 \ee
and thus the fiducial temperature is
 \be\label{unruh-e}
 T_{\rm FID}=\frac{1+\frac{Q^2}{Mr}}{8\pi M\sqrt{1+\frac{Q^2-2M^2}{Mr}}}.
 \ee

On the other hand, for the case of the electrically charged
solution (\ref{metric-s2}), we have orbit equations as
 \ba
 \frac{dt}{d\tau}&=&  \frac{\left(1+\frac{Q^2-2M^2}{Mr_0}\right)^{1/2}}{\left(1+\frac{Q^2}{Mr_0}\right)}
                     \frac{\left(1+\frac{Q^2}{Mr}\right)^2}{\left(1+\frac{Q^2-2M^2}{Mr}\right)},
                    \nonumber\\
 \frac{dr}{d\tau}&=&-\left[-\left(1+\frac{Q^2-2M^2}{Mr}\right)
                     +\left(1+\frac{Q^2}{Mr}\right)^2
                     \frac{\left(1+\frac{Q^2-2M^2}{Mr_0}\right)}{\left(1+\frac{Q^2}{Mr_0}\right)^2}\right]^{1/2},\nonumber\\
 \ea
which can be used to get the $\widetilde{a}_7$ acceleration for a
freely falling observer given by
 \be\label{local-a7-ele}
 (\widetilde{a}_7)^2=\frac{h_e(r,M,Q)}{16M^2\left(1+\frac{Q^2}{Mr}\right)^2},
 \ee
where
 \ba
  h_e(r,M,Q)= \left[\left(1+\frac{2M}{r}+\frac{4M^2}{r^2}+\frac{8M^3}{r^3}\right)
                +\frac{Q^2}{2M}\left(3+\frac{4M}{r}+\frac{4M^2}{r^2}\right)\right. \nonumber\\
            +\left.\frac{Q^4}{4M^2}\left(3+\frac{2M}{r}+\frac{8M^3}{r^3}+\frac{Q^2}{Mr}\right)\right]
              -\frac{4M^2 Q^2}{r^3}\left(2+\frac{Q^2}{2M^2}+\frac{Q^4}{2Mr^3}\right).\nonumber\\
 \ea
As a result, we have the local free-fall temperature for a freely
falling observer at rest as
 \be\label{ffar-ele}
 T_{FFAR}=\frac{\widetilde{a}_7}{2\pi}
 \ee
At the asymptotic infinity the local free-fall temperature again
becomes the Hawking temperature $T_H$. Note that when $Q\rightarrow
0$, the local $FFAR$ temperature (\ref{ffar-ele}) reduces exactly
again to the local free-fall temperature (\ref{ffar-sch}) for the
Schwarzschild spacetime. At the event horizon, the local free-fall
temperature also remains finite as
 \be
  T_{FFAR}=\frac{\sqrt{1-\frac{3Q^2}{4M^2}}}{4\pi M\left(1-\frac{Q^2}{2M^2}\right)},
 \ee
while the fiducial temperature (\ref{unruh-e}) diverges.

On the other hand, in the extremal limit of $2M^2=Q^2$, it becomes
 \be
 T_{FFAR}=\frac{1}{8\pi M}\frac{\sqrt{1+\frac{8M}{r}+\frac{24M^2}{r^2}
                                   +\frac{32M^3}{r^3}-\frac{48M^4}{r^4}}}
                               {\left(1+\frac{2M}{r}\right)}.
 \ee
This contrasts with the magnetically charged extremal $T_{FFAR}$
in Eq.~(\ref{extT-mag}), which is independent of $r$, while the
electrically charged extremal $T_{FFAR}$ is a function of $r$.
However, as $r\rightarrow\infty$, the temperature $T_{FFAR}$
becomes the same as the local free-fall temperature
$T_{FFAR}$~(\ref{extT-mag}) for the magnetically charge solution.

\begin{figure}[t!]
   \centering
   \epsfbox{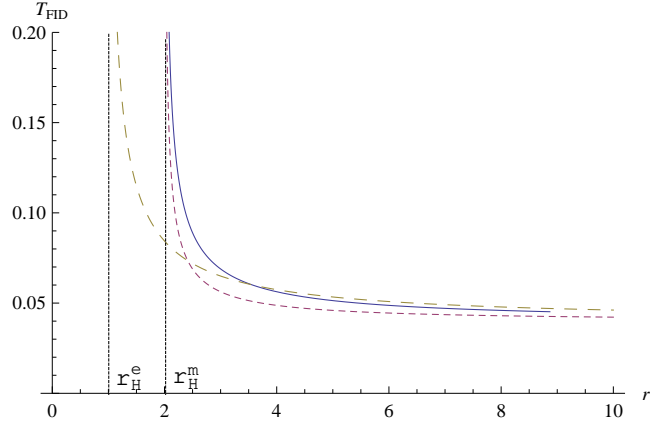}
\caption{The local temperatures $T_{\rm FID}$ measured by a fiducial
observer for the nonextremal GMGHS spacetime with $M=1$, $Q=1$: the
solid line for the solution in Einstein frame, the dotted line for
magnetically charged solution in string frame, and the dashed line
for electrically charged solution in string frame.} \label{fig1}
\end{figure}

\begin{figure}[t!]
   \centering
   \epsfbox{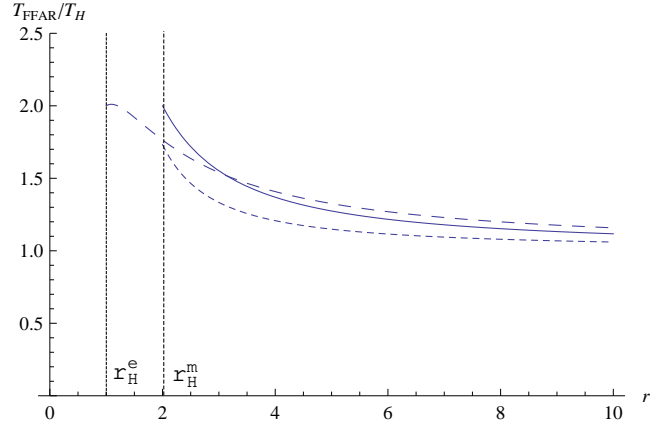}
\caption{The local temperatures $T_{\rm FFAR}$ and $T_{\rm FID}$ for
the nonextremal GMGHS spacetime with $M=1$, $Q=1$: the solid line
for the solution in Einstein frame, the dotted line for magnetically
charged solution in string frame, and the dashed line for
electrically charged solution in string frame.} \label{fig2}
\end{figure}

In short, we have plotted in Fig.~\ref{fig1} the local
temperatures $T_{\rm FID}$ measured by the fiducial observers for
the nonextremal GMGHS spacetime. All the fiducial temperatures
show the same behaviors that they blow up near the event horizons
and become the Hawking temperatures at asymptotic infinities. On
the other hand, Fig.~\ref{fig2} shows that the local free-fall
temperatures obtained in the string frames as well as in Einstein
Einstein frame remain finite at the event horizons. Note that in
the Einstein frame, the temperature is independent of the charge,
while in the string frames the temperature depends on the charge
so that the temperature differs between the two frames.

%%%%%%%%%%%%%%%%%%%%%%%%%%%%%%%%%%%%%%%%%%%%%%%%%%%%%%%%%%%%%%%%%%%%%%%
\section{Conclusions}
\setcounter{equation}{0}
\renewcommand{\theequation}{\arabic{section}.\arabic{equation}}
\label{sec:conclusions}
%%%%%%%%%%%%%%%%%%%%%%%%%%%%%%%%%%%%%%%%%%%%%%%%%%%%%%%%%%%%%%%%%%%%%%\vskip10pt

In this paper, we have obtained the (5+1)/(5+2)-dimensional global
flat embeddings of the GMGHS spacetime according to the
Einstein/string frames. In the Einstein frame, we need the
(5+1)-dimensional embedding, which is similar to the embedding of
the Schwarzschild spacetime. In some sense, it is expected since the
metric is identical to the Schwarzschild black hole metric in the
$(t-r)$ plane and thus the casual structure is the same. However, by
changing the Einstein frame to the string frame, we have shown that
regardless of the type of the charges the global flat embeddings of
the GMGHS spacetime are needed one more time dimension as like the
RN spacetime. In this respect, the GEMS of the GMGHS spacetime in
the string frame is more like the ones of the RN spacetime.

We have also found the Unruh/fiducial temperatures in each frame
and their corresponding Hawking temperatures. Moreover, by finding
local free-fall temperatures for the freely falling observers, we
have shown that the local free-fall temperature in the Einstein
frame as well as in the string frame remains finite at the event
horizon, while the Unruh/fiducial temperature is divergent. On the
other hand, regardless of the frames, all the temperatures are
reduced to the Hawking temperature $T_H$ at infinity.

\section*{Acknowledgement}
This work was supported by the National Research Foundation of Korea
(NRF) grant funded by the Korea government (MISP) through the Center
for Quantum Spacetime (CQUeST) of Sogang University with grant
number 2005-0049409.

\end{document}